TYPE Review
PUBLISHED 27 July 2023
DOI 10.3389/feduc.2023.1210968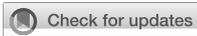

OPEN ACCESS

EDITED BY
Olga Scrivner,
Rose-Hulman Institute of Technology,
United States

REVIEWED BY
Myungjae Kwak,
Middle Georgia State University, United States
Alina Bockshecker,
University of Hagen, Germany

*CORRESPONDENCE
Hasan Abu-Rasheed
✉ hasan.abu.rasheed@uni-siegen.de

RECEIVED 23 April 2023
ACCEPTED 04 July 2023
PUBLISHED 27 July 2023

CITATION
Abu-Rasheed H, Weber C and Fathi M (2023) Context based learning: a survey of contextual indicators for personalized and adaptive learning recommendations — a pedagogical and technical perspective.
Front. Educ. 8:1210968.
doi: 10.3389/feduc.2023.1210968COPYRIGHT
© 2023 Abu-Rasheed, Weber and Fathi. This is an open-access article distributed under the terms of the Creative Commons Attribution License (CC BY). The use, distribution or reproduction in other forums is permitted, provided the original author(s) and the copyright owner(s) are credited and that the original publication in this journal is cited, in accordance with accepted academic practice. No use, distribution or reproduction is permitted which does not comply with these terms.# Context based learning: a survey of contextual indicators for personalized and adaptive learning recommendations – a pedagogical and technical perspective

Hasan Abu-Rasheed*, Christian Weber and Madjid Fathi

Department of Electrical Engineering and Computer Science, Institute of Knowledge Based Systems and Knowledge Management, University of Siegen, Siegen, GermanyLearning personalization has proven its effectiveness in enhancing learner performance. Therefore, modern digital learning platforms have been increasingly depending on recommendation systems to offer learners personalized suggestions of learning materials. Learners can utilize those recommendations to acquire certain skills for the labor market or for their formal education. Personalization can be based on several factors, such as personal preference, social connections or learning context. In an educational environment, the learning context plays an important role in generating sound recommendations, which not only fulfill the preferences of the learner, but also correspond to the pedagogical goals of the learning process. This is because a learning context describes the actual situation of the learner at the moment of requesting a learning recommendation. It provides information about the learner's current state of knowledge, goal orientation, motivation, needs, available time, and other factors that reflect their status and may influence how learning recommendations are perceived and utilized. Context-aware recommender systems have the potential to reflect the logic that a learning expert may follow in recommending materials to students with respect to their status and needs. During the last decade, several approaches have emerged in the literature to define the learning context and the factors that may capture it. Those approaches led to different definitions of contextualized learner-profiles. In this paper, we review the state-of-the-art approaches for defining a user's learning-context. We provide an overview of the definitions available, as well as the different factors that are considered when defining a context. Moreover, we further investigate the links between those factors and their pedagogical foundations in learning theories. We aim to provide a comprehensive understanding of contextualized learning from both pedagogical and technical points of view. By combining those two viewpoints, we aim to bridge a gap between both domains, in terms of contextualizing learning recommendations.

# 1. Introduction

In modern educational practices, whether in formal education, vocational training, digital learning platforms or otherwise, the strategy of one-size-fits-all has proven insufficient (Siraj et al., 2018). Even learners who have the same learning goal still come from different backgrounds, have different learning practices and preferences, and learn in different ways. Personalizing the learning experience for individuals or groups of learners is needed to ensure a higher learning outcome (Taylor et al., 2021), more engagement and satisfaction (Rajabalee and Santally, 2021) and life-long learning commitment (Ramírez Luelmo et al., 2021). Personalizing the learning experience takes different forms that include, but are not limited to, the selection of learning materials for individual learners, the order of steps on learning paths toward a learning goal, the conditions of the learning, such as time and duration, as well as the means of learning, such as using certain digital tools or accessing different types of learning materials.

While educators rely on their knowledge and experience to personalize the learning for their students, digital learning platforms, and technology enhanced learning (TEL) approaches in general, depend on filtering and recommendation algorithms to achieve the goal of finding personalized content for a learner, or creating personalized learning paths (Manouselis et al., 2013). This has been one part of the overall development of personalized and adaptive learning environments (PALE), which builds on the pedagogical foundations for learning and utilizes available technologies to support their implementation (Taylor et al., 2021). Recommendation systems in TEL and PALE use the information available about the learner and the learning content to generate a personalized recommendation. Since learning always takes place within a certain context (Brown et al., 1989), information about that context has been considered in a variety of recommendation systems, namely context-aware recommendation systems, to personalize the learning recommendations not only based on learner preferences, but also depending on the context, in which learning happens.

The definition of "context," however, varied among researchers in the literature, as reviewed by Verbert et al. (2012). Differences of those definitions played a role in defining different dimensions and factors, which are considered by the recommender system to "capture" the learning context. For example, a definition of the context from the perspective of context-aware recommender systems indicates that it is an aggregation of multiple categories, which describe the setting in which the recommender is implemented (Verbert et al., 2012). Those categories can be the location, nearby learners, or the level of noise around the learner. On the other hand, a context definition from a pedagogical point of view, as in Koubek et al. (2009), describes the context as the dimensions of the situations, which a student can experience. Those not only include the technology used or the demographic information, but also the ethical aspects, law, or economy. Despite the overlap between the two definitions, the differences reflect a certain focus from each domain on the factors that describe the learning context of a student. In digital and e-learning platforms, pedagogical foundations from learning theories should be considered by recommendation system developers to achieve the pedagogical goals of learning. This means that the definition of context and the indication of its factors should be consistent, and aligned with pedagogical requirements and considerations of the learning context (Isaias et al., 2022).

The objective of this article is to provide a comprehensive summary of the existing literature on the contextual factors used within the domain of TEL and in the light of the pedagogical foundation from learning theories. We review the different definitions of the learning context, which have been adopted in the literature during the 11-year period between (2012–2022), within the domains of TEL and context-aware recommendation systems. We integrate the pedagogical foundation of those definitions, by further investigating the context definitions in different learning theories. We then survey the context factors that have been used by different researchers from both points of view (pedagogical and technical) and highlight the similarities and differences in the use of contextual factors, as well as the indications of those factors.

Based on the scope of this study and the above-mentioned objectives, we formulate the research questions through a preliminary review of the existing literature and similar surveys on the topic. This allowed identifying the knowledge gaps and areas that required further exploration. Research questions were formulated based on these findings, focusing on the following aspect:

- What contextual factors are used in the literature in the fields of TEL and context-aware recommendations?
- What are the categories, to which those factors belong?
- Which factors are considered based on a pedagogical foundation from learning theories?
- Which learning theories are addressed?
- What sources of contextual information are considered? (i.e., user, learning material, or learning environment context).

In the following sections of this paper, and following the systematic literature review process in Vom Brocke et al. (2009), we present the conceptualization background of this review in section 2, along with highlighting similar reviews in the same domain. We then describe the research process and review methodology in section 3. Analysis of the surveyed literature is presented in section 4, while we synthesize the results, patterns and observations in section 5. The paper is then concluded in section 6, with a highlight on the limitations and future work.

# 2. Background and related work

Learning theories aim to provide a framework for understanding how learning occurs and how it can be facilitated (Mitchell and Govias, 2021). They help to understand the cognitive, emotional, and behavioral processes involved in learning, as well as the factors that influence learning outcomes. A solid understanding of the learning context, and how it could be captured and implemented in TEL and PALE based on concrete pedagogical foundations, requires an understanding of how learning theories, TEL, and PALE are defined, and how they address the context in their philosophies (Isaias et al., 2022). In this section, we explore the most influential learning theories and their implications for education and psychology, as well as the main features and characteristics of TEL and PALE.

## 2.1. Learning theories and approaches

A learning theory explains how people learn, identifies factors that influence learning, and provides guidance on how to enhance learning outcomes. It provides a framework for educators to design effective teaching and learning environments. Learning theories have evolved over time as new research, technologies and perspectives emerge. Some of the most influential learning theories include behaviorism, cognitivism, constructivism, and connectivism (Ertmer and Newby, 2013). Each theory emphasizes different aspects of learning, such as the role of external rewards and punishment, mental processes involved in learning, the importance of learners' experiences and context, and the role of technology in learning. By understanding and applying learning theories, educators can design personalized and effective learning experiences that meet the needs of diverse learners. We introduce in the following sub-sections the main learning theories that were addressed in the reviewed literature.

### 2.1.1. Behaviorism

Behaviorism (Skinner, 1953) is a learning theory that emphasizes the role of the environment in shaping behavior. According to this theory, learning is the result of the interaction between the individual and the environment, where behavior is shaped by the consequences that follow it. Behaviorists believe that behaviors can be reinforced or punished, and that learning occurs when individuals associate a particular behavior with a positive or negative outcome. Thus, behaviorism focuses on observable behaviors and measurable outcomes, and it has been used to develop effective teaching strategies based on rewards and punishments.

### 2.1.2. Cognitivism

Cognitivism (Piaget and Cook, 1952) is a learning theory that emphasizes the role of mental processes in learning. According to this theory, learning is the result of the interaction between the individual's cognitive processes and the environment. Cognitivists believe that individuals actively construct knowledge and make sense of their experiences through cognitive processes such as attention, perception, memory, and problem-solving. Thus, the cognitivist theory focuses on the mental processes that occur during learning, and it has been used to develop effective teaching strategies based on active engagement and meaningful learning (Ausubel, 1963).

### 2.1.3. Connectivism

Connectivism (Siemens, 2004) is a learning theory that emphasizes the role of technology and networks in learning. According to this theory, learning is an interconnected process that occurs through networks of people, information, and technology. Connectivists believe that learners can access and share information through online networks, and that this can lead to the creation of new knowledge and the development of new skills. Thus, the connectivist theory focuses on the use of technology, as well as information and social networks to facilitate learning, and it has been used to develop effective teaching strategies based on networked learning and digital literacy.

### 2.1.4. Constructivism

Constructivism (Vygotsky and Cole, 1978; Von Glasersfeld, 2013) is a learning theory that emphasizes the role of the learner in constructing knowledge. According to this theory, learning is an active process of constructing meaning from new information and experiences. Constructivists believe that learners actively create their own understanding of the world by building upon their prior knowledge and experiences. Thus, the constructivist theory focuses on the learner's active engagement in the learning process, and it has been used to develop effective teaching strategies based on problem-solving, discovery learning, and collaborative learning.

*Social constructivism* (Vygotsky and Cole, 1978), an extension of the constructivist theory, emphasizes the social and cultural aspects of learning. According to social constructivism, learning is not only an individual process, but also a social one. Social interactions and collaborative activities help learners construct new knowledge and understandings. This theory suggests that knowledge is co-constructed through social interactions, and that learning takes place within a social context. Social constructivism highlights the importance of social and cultural factors in shaping individuals' knowledge and understanding.

*Radical constructivism* (Von Glasersfeld, 2013), another variant of the constructivist theory, argues that knowledge is constructed by individuals, rather than discovered or acquired from the environment. This theory suggests that knowledge is subjective and that it can only be understood from the perspective of the individual. Radical constructivism emphasizes the importance of personal experience and interpretation in constructing knowledge. According to this theory, there is no objective reality that can be directly known, but rather only individual interpretations of reality.

### 2.1.5. Situated learning

Situated learning (Lave and Wenger, 1991) proposes that learning is not just the acquisition of knowledge and skills, but also the social and cultural practices in which learning occurs. In situated learning, learning is seen as a social activity that is situated in the real-world environment, rather than a process that occurs solely within the learner's mind. Situated learning suggests that knowledge and skills are closely tied to the situations in which they are acquired and used, and that learning is more effective when it is situated in real-world contexts.

### 2.1.6. Meaningful learning

Meaningful learning (Ausubel, 1963) suggests that learners construct knowledge by integrating new information and experiences into their existing knowledge structures. Meaningful learning occurs when learners actively engage with new information, make connections to their prior knowledge, and apply their understanding to new contexts. The goal of meaningful learning is for learners to acquire a deep and flexible understanding of the material, rather than just memorizing facts or procedures.

### 2.1.7. Self-determination theory

This theory (Deci and Ryan, 1985) proposes that learners are motivated to learn when they feel a sense of autonomy, competence, and relatedness. Autonomy refers to the learner's sense of control over their own learning process, competence refers to the learner's sense of ability to achieve their goals, and relatedness refers to the learner's sense of connection to others. When learners feel these three needs are being met, they are more likely to engage in self-regulated learning and persist in the face of challenges.

Derived from the previous theories and related to them are several pedagogical approaches, instructional design models or motivational theories, which were also clearly present in the reviewed literature:

### 2.1.8. Student-centered learning

It is also referred to as learner-centered pedagogy, and it is grounded on the principles of constructivist learning theory. It empowers students to take charge of their learning by enabling them to make informed decisions in the learning process (Hannafin and Land, 1997)–(Goodman et al., 2018). As Dockterman (2018) posits out that students learn more effectively when instruction is tailored to their individual needs, interests, and skills. Personalized learning acknowledges the diversity of students, and this has led to the emergence of a new pedagogy of personalization. However, to implement personalized learning on a large scale, there is a need for technological intervention, which has been lacking until recently. Adaptive learning platforms have been developed to identify learners' needs and offer appropriate support for effective learning (Taylor et al., 2021).

### 2.1.9. Scaffolding learning

Scaffolding is defined as the support and guidance provided to the learner until the learner can accomplish a task or demonstrate competence independently (Wood et al., 1976). This theory proposes that learners can accomplish more with the help of a knowledgeable and skilled mentor. Scaffolding involves providing support, guidance, and feedback to learners as they engage in challenging tasks. The mentor's role is to gradually reduce the amount of support as the learner becomes more capable, until the learner is able to complete the task independently. The goal of scaffolding is to help learners achieve a level of competence that would require otherwise more added effort to reach on their own.

### 2.1.10. Problem-based learning

Problem-based learning (PBL) is a student-centered instructional approach that emphasizes the active engagement of students in solving ill-structured problems (Savery and Duffy, 1995). In PBL, students work collaboratively in small groups to explore complex problems, identify gaps in their knowledge, and develop solutions. PBL is grounded in constructivist learning theory, which emphasizes the active construction of knowledge through hands-on, experiential learning (Jonassen, 1991). By working on real-world problems, PBL encourages learners to integrate and apply their knowledge in meaningful ways, rather than simply memorizing isolated facts.

### 2.1.11. Goal orientation theory

Goal orientation (Ames, 1992; Elliot and McGregor, 2001) suggests that learners' motivation and learning behavior are influenced by their goals. There are two main types of goals: mastery goals and performance goals. Mastery goals focus on learning and improving one's abilities, while performance goals focus on demonstrating competence and outperforming others. Learners who adopt mastery goals are more likely to engage in deep learning strategies and persist in the face of challenges, while learners who adopt performance goals are more likely to engage in surface learning strategies and give up when faced with obstacles.

## 2.2. Contextualized learning within the intersection of learning theories

Learning theories do overlap, being informed by the same or related disciplines, as well as commonly seek to explain how learners acquire new knowledge, skills, and behaviors (Mitchell and Govias, 2021). Situated learning, meaningful learning, and scaffolding all emphasize the importance of active engagement in the learning process, while goal orientation theory and self-determination theory focus on the role of motivation in learning. Together, these theories provide a comprehensive framework for understanding the complex process of learning.

It is also clear from the definition of theories that several of them highlight the importance of the learner's context in the learning process, including:

1. Situated learning theory emphasizes the importance of the social and physical context in which learning occurs (Lave and Wenger, 1991).
2. Social constructivism theory posits that knowledge is constructed through social interaction and that learning is influenced by the learner's cultural and social background (Agarkar and Brock, 2017).
3. Constructivist theory emphasizes the importance of the learner's prior knowledge and experiences, which are shaped by their personal and social context (Knobelsdorf and Tenenberg, 2013).
4. Self-determination theory emphasizes the importance of creating a learning environment that supports the learners' autonomy, relatedness, and competence needs (Deci and Ryan, 1985).

Inquiry-based learning and project-based learning are also approaches that can be applied within the context of these theories, and they involve creating a learning environment that is tailored to the learner's needs and interests. Those approaches, however, were not in the focus of the reviewed literature, so we leave it to the reader to investigate more on their definition and applications.

## 2.3. Technology enhanced learning and personalized and adaptive learning

TEL is an approach that integrates technological tools to facilitate and enhance the learning process (Manouselis et al., 2013). TEL provides opportunities for learners to access and interact with digital resources, collaborate with peers and instructors, and receive feedback and support in a timely and personalized manner. It involves the integration of various digital tools and resources, such as multimedia content, online discussion forums, and mobile applications, into educational settings to facilitate effective learning and knowledge transfer (Anderson, 2008). TEL has gained popularity in recent years due to its potential to increase access to education, improve learner engagement and motivation, and provide personalized learning experiences. Moreover, TEL can provide opportunities for learners to collaborate and engage with their peers and instructors, even if they are not physically present in the same location. This can be achieved through various online communication tools, such as video

conferencing, social media, and virtual learning environments (Siemens and Baker, 2012). These tools enable group discussions, peer feedback, and collaborative project work, which can enhance learners' critical thinking and problem-solving skills.

TEL can be used to support various learning theories, such as constructivism, connectivism, and social learning. For example, TEL can provide learners with access to a wide range of resources and tools, enabling them to construct their own knowledge through exploration and experimentation. TEL also aligns with social constructivism by facilitating social interactions and knowledge-sharing among learners, which can promote the co-construction of knowledge. Additionally, TEL can support connectivism by providing learners with access to a wealth of resources and networks, which can help them build their own personal learning networks.

One of the key benefits of TEL is its potential to enable personalized and adaptive learning environments. Personalized and adaptive learning (PAL) is a pedagogical approach that focuses on tailoring learning experiences to meet the unique needs, interests, abilities, and preferences of individual learners by adapting the learning content, pace, and strategies to these learners' characteristics. PAL uses a variety of data sources, including performance metrics, learner feedback, and demographic information, to generate customized learning experiences that are designed to optimize learning outcomes. The goal of PAL is to improve learners' engagement, motivation, and achievement by providing them with tailored content and experiences that match their unique learning styles and preferences.

PAL analyzes learner data and generates personalized content and activities. For example, an adaptive learning system may use data on a learner's past performance and knowledge gaps to generate targeted exercises and quizzes that focus on areas where the learner needs additional support (Johnson et al., 2015). Such systems may also adjust the difficulty level of content and activities based on learners' performance to ensure that they are appropriately challenged and engaged (Moltudal et al., 2022).

Another important feature of PAL is its focus on learner agency and control. PAL systems often provide learners with a range of options and choices, allowing them to decide what and how they want to learn. This can include options for pacing, content selection, and learning activities (Dabbagh and Kitsantas, 2012). By giving learners more control over their learning experience, PAL systems can help to increase motivation and engagement and promote a more student-centered learning approach.

TEL provides a range of tools and resources that can facilitate PALE, including intelligent tutoring systems, learning analytics, educational games, and virtual simulations. For example, TEL can be used to provide learners with customized learning paths, adaptive feedback, and personalized resources based on their learning preferences. TEL can also be used to monitor learners' progress and provide timely interventions to support their learning.

TEL is relevant to a wide range of learning theories, including constructivism, connectivism, and social learning. TEL can provide learners with access to a wealth of resources and tools, enabling them to construct their own knowledge through exploration and experimentation. TEL can also facilitate social learning by enabling learners to collaborate and communicate with others, thereby enhancing their understanding and problem-solving skills. Additionally, TEL can be used to support the development of metacognitive and self-regulated learning skills, as learners can use technological tools to monitor and reflect on their learning progress.

PALE draws on learning theories such as self-determination theory, which highlights the importance of learners' autonomy, competence, and relatedness in motivation and engagement, and situated learning theory, which emphasizes the role of the context and social interaction in learning.

## 2.4. Context definitions between pedagogy and technology

According to Merriam-Webster dictionary, the term "context" is lingually defined as "*the interrelated conditions in which something exists or occurs*" (Merriam-Webster, 2023). Originally, the term described the part of discourse, which surrounds a word (Merriam-Webster, 2023). This definition was later extended to include events in addition to words. In this sense, the context refers to the environment or setting in which something exists. Following this notion, a "contextualized" entity is one that is "*placed in an appropriate setting, one in which it may be properly considered*" (Merriam-Webster, 2023).

The literal definition of context is then interpreted and implemented differently, depending on the domain and the nature of the entity itself. For example, in pedagogy, a learner's context can describe the situation of that learner, in terms of their social connections (Vygotsky and Cole, 1978) motivation (Ames, 1992) or current level of knowledge (Kuger and Klieme, 2016), while a technical consideration of the same learner's context can be described with the sensory data that a mobile device collects from the learner's interaction with the learning material on the device (Alnuaim et al., 2016). In the literature, multiple definitions of the learning context can be identified, which are mainly influenced by their dependency on pedagogical foundations or technical ones. The overlap and differences between those definitions are not only on the wording level of the definition itself, but also on the implementation of context-capturing approaches in a real-world scenario. In an interdisciplinary domain that addresses both pedagogical and technical requirements of the learning process, such as TEL, a comprehensive and unified approach to defining and utilizing learning context is essential, yet it is still rarely investigated in the literature (Mayeku and Hogrefe, 2017). When it comes to technical implementations of the context in a learning process, the meaning of contextual concepts like "time," "space" or "place" has been under-theorized (Pimmer et al., 2013). This, in turn, reflects on the approaches used for capturing these concepts, which will also not be theory- or pedagogy-driven, but rather technology-driven. In that case, the real pedagogical value of the resulting contextualization methods cannot be guaranteed.

In situated learning theory, the definitions of learning context extend from the learner themselves to include their "interactions" with the environment (Bredo, 1994; Hung and Chen, 2001). In this sense, knowledge is not an exclusive element in the learner's mind, but it is rather situated as a part of the activity or culture, where it has been developed, or where it is used (Pimmer et al., 2013).

A "meaningful context" in the field of vocational education and training is not separable from its socialization aspect, which refers to the cooperative character of work processes and their influence on the division of labor, e.g., to lead to cost reduction or digitization (Buchmann, 2022). This aspect of the context shows that other dimensions, such as the legal and psychological ones, are to be considered in addition to the pedagogical ones in defining the context. This definition has its origins in social constructivism and connectivism, where the role of social interaction and the technologies that enable it is key in creating the knowledge in the mind of the learner.

In contrast to the philosophical nature of context definitions in pedagogy, and its focus on knowledge construction in the learner's mind, context awareness in technology has been defined by some scholars as the ability of software and hardware to predict and propose services to a user based on analyzing their position, time, living patterns, surrounding space, bodily signals and vital signs (Ku, 2014). Learning contextualization in this sense refers to using sensors, e.g., of a mobile device, to collect information about the learner to personalize and adjust the learning offers they receive to their current context.

While the two previous definitions build on the same merits of the general context definition, being the situation in which learning happens, it is clear that the focus of each definition is different. It is important, however, to point out here that contextual definitions are not opposites between pedagogy and technology, but rather spread over a spectrum of slightly different interpretations of "what the learning situation is" and "how it can be captured." Between the two fields, PALE and TEL research has provided context definitions that address both technical and pedagogical requirements for contextual learning and personalization. A common context definition in this field is proposed by Dey (2001) which addresses "any information that can be used to characterize the situation of an entity," where the entity can then be a person, place, or object. The information in this definition refers to any element or piece of data that enables describing the condition or state of the entity (Gómez et al., 2014). In TEL particularly, Luckin defines the context as "the current situation of a person related to a learning activity" (Luckin, 2010).

## 2.5. Related work

During the reviewed period, there has been a sum of surveys on context-aware recommendation systems. This is an understandable observation, due to the growing attention to this type of recommendation systems, both commercially and pedagogically. While surveys focused on recommendation algorithms and implementations, the number of surveys that addressed the contextual factors considered by such recommender systems, especially in TEL and PALE domains, was limited in the state of the art, see Table 1. This especially applies when furthermore considering the pedagogical foundations of the context. After 2012, we identify a gap in literature reviews focusing on context definitions and context factors, which are informed by pedagogy and learning theories.

In 2012, Verbert et al. (2012) have thoroughly investigated the context of recommender systems in the TEL domain. Authors introduced a context framework for this type of recommenders, which identifies a set of context dimensions. Their survey analyzes the definitions of learning context in TEL from pedagogical and technical points of view, as well as the variety of factors used to capture that context. Thus, the authors offer a comprehensive understanding of the contextual factors in TEL. In 2013, Hwang (2013) investigated context-aware ubiquitous learning, highlighting the role of the pedagogical and theoretical concepts of e-learning in shaping the definition and factors of the ubiquitous learning context.

Following those two surveys, we find the next survey that focuses on the pedagogical implications of defining the contextual factors in the work of Hemmler and Ifenthaler (2022). The authors revisit the role of empirical-pedagogical research in shaping the design of PALE and in generating meaningful, context-aware, recommendations. Their survey identifies 208 internal and external contextual factors, spread over 27 dimensions. The survey also points out the lack of literature reviews, which focus on the pedagogical aspects of the learning context and its factors. While the authors put a focus on context indicators in higher-education, workplace learning, learning analytics, competence acquisition and competence assessment, as indicated by the search terms adopted in their survey, the researchers address that the time span covered in the survey was limited, going back only to 2019. The authors also do not thoroughly discuss the definitions of the context and the indicators identified. Another recent work by Ahmad et al. (2022) has introduced a repository of indicators, along with a dashboard tool to support educators in the selection of learning activities. Although the authors mainly propose the tool and evaluate it in their work, they still build the selection of the indicators on a review of the literature, which they conducted between 2011 and 2021. The authors do not discuss, however, the indicators themselves, their definitions, or their pedagogical and technical foundations, in their work.

We, therefore, build on the important studies carried out by Verbert et al. and Hemmler and Ifenthaler and extend the survey of contextual factors in TEL in the light of pedagogical and technical considerations of the learning process. We highlight in this survey the definitions and dimensions of learning context in the period between 2012 and 2022, thus extending the work of Verbert et al. We also address the learning theories that were the bases for the pedagogical foundation of defining the context in TEL and the selection of the indicators and factors to capture that context. We focus on both the technical factors and those resulting from pedagogy since both are essential to support a pedagogically informed digital learning recommendation. Therefore, the search terms we utilize for literature search and data collection in this survey are tailored directly to the contextual factors and dimensions in TEL and learning theories.

## 3. Methodology

In order to achieve the objectives of the survey, a systematic literature review methodology was employed, following the framework and guidelines proposed by Vom Brocke et al. (2009) and

TABLE 1 Related reviews in the literature within the review period.

| Review | Year | Reviewed period | Focus |
| --- | --- | --- | --- |
| Verbert et al. (2012) | 2012 | – | TEL, recommender systems |
| Hwang (2013) | 2013 | – | Ubiquitous learning |
| Hemmler and Ifenthaler (2022) | 2022 | 2007–2021 | PALE, Pedagogy |
| Our | 2023 | 2012–2022 | Pedagogy, TEL and recommender systems |

Webster and Watson (2002). After defining the scope and objective of the survey in the introduction, and conceptualizing the topic in the Background section, we define here the literature search methodology. We then follow up with analyzing the literature and synthesizing the results of the survey. By the end of the survey, we provide the reader with observations made from the analysis and synthesis phase, which is meant to represent a form of research agenda for potential topics the literature may consider focusing on in the future.

## 3.1. Literature search strategy and data collection

The search strategy was developed to ensure, to the best of our knowledge, that all relevant literature on the topic was identified and collected. A comprehensive search was conducted using various databases, namely Google Scholar, ACM Digital Library, Springer Link, EbscoHost, and LearnTechLib. The search was limited to the period between 2012 and 2022. The databases and search engines were selected based on their relevance to the topic and their comprehensiveness, in terms of covering high-quality literature on the technical and pedagogical domains as sources of the literature on learning context and its factors.

The search terms were identified based on the research questions and developed using a combination of keywords and controlled vocabulary. They were selected to ensure that all relevant literature on the topic is retrieved, and to minimize the risk of missing important studies. Selected search terms are:

- *[(learner OR learning) AND context] AND [TEL OR e-learning OR (technology AND enhanced AND learning)]*
- *[(learner OR learning) AND context] AND (learning AND theories)*
- *[(learner OR learning) AND context] AND (indicators OR factors OR dimensions)*
- *[(learner OR learning) AND context] AND (modeling OR model)*

Each of the selected databases has its own search engine. To accommodate those terms in the search engine of each database, the search strategy was adjusted to follow the technical requirements of that engine, assuring that the results generated from searching the different databases are based on the same search-term-combination logic.

Following the taxonomy of literature reviews introduced by Cooper (1988), who introduced six characteristics of the review, ours falls into the following categories: (1) Focus: we focus on the theories that lie behind the definition, selection and utilization of context factors. (2) Goal: we aim for the integration of concepts of contextualization between pedagogy and technology. (3) Organization: we try to organize the review based on the concepts and methods introduced in the literature rather than its historical order. (4) Perspective: we seek an objective observation of the literature and a neutral analysis from the differences and overlaps of the domains. (5) Audience: specialized scholars in the domains of pedagogy, TEL, and PALE, are targeted in this review. (6) Coverage: we follow an exhaustive and selective approach, through seeking a comprehensive search for the literature in the field and applying selection criteria to limit the scope of the survey within the boundaries of its objectives.

After retrieving the search results from all selected databases, we analyze the most frequent terms in their titles and abstracts, to control for the alignment with the research objectives. Figure 1 shows word cloud visualization of the most common terminology in the retrieved results.

## 3.2. Selection criteria

The articles were screened based on their relevance to the research question, inclusion and exclusion criteria, and quality assessment.

Inclusion criteria for the review are that the papers selected have to:

- represent original work or be a survey
- be accessible to the authors of this study
- be written in the English language

Exclusion criteria were defined to eliminate papers that appear in the search results but do not serve the goal of the review. These criteria are:

- The paper considers a context that is not a "learning context."
- The paper is a workshop description, proceedings' preface, or similar.
- The paper was already found in another database (duplicated).

In addition to the retrieved papers from the databases, a group of eight papers was added to the study. Those papers were not among the search results since the search terms do not appear in their titles or abstracts. They were, however, identified as relevant due to their content that implicitly describes contextual factors. Additional papers were found either in the references of papers included in the review, or based on the authors' previous knowledge about their content.

## 3.3. Literature screening process

Reviewers screened the papers based on their titles and abstracts. When there was doubt about the inclusion or exclusion of a paper, the full-text of that paper was then screened to identify the inclusion and exclusion criteria fulfillment. The final papers considered in the review were then analyzed to extract findings relevant to the research questions. Figure 2 illustrates the screening and selection process of the papers included in this survey.

# 4. Analysis of the surveyed literature

In this section, we discuss the analysis of the surveyed literature and highlight the findings form the concepts identified. To perform this analysis, data collected in the previous phase has been arranged in a database, which represented each paper with the concepts it included. We use Zotero literature management tool to create a report of each paper, reflecting the concept it includes regarding the research questions of this survey. We, investigate and document what contextual factors are addressed in the paper, the categories and sources of those factors, the application domain(s) the paper covers,

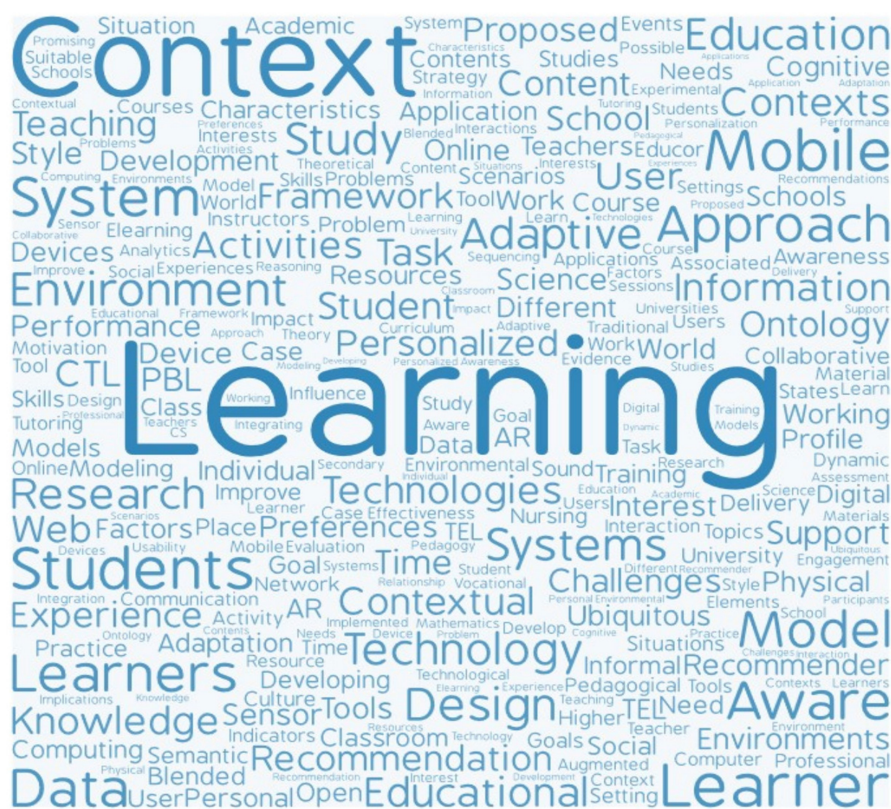

FIGURE 1
Most common terms appearing in titles and abstracts of retrieved search results.

the foundation of the factor selection and the learning theories addressed, if any.

The overall number of papers reviewed in this research is spread over the period between 2012 and 2022. In the selection process, papers that considered the "context" were nominated for the final screening phase. From those papers, we have found that not all authors explicitly addressed the "learning context," which is our focus in this survey. Rather than an explicit learning context, these papers described context-based algorithms or approaches which could be implemented in a learning-related application. This portion of papers still addressed contextual factors and dimensions, but from a general point of view that does not necessarily apply to a learner. We point out those papers to highlight the tendency and patterns in the literature for considering a "learner context" in the learning process" against a "user context" in a general application. Figure 3 shows the number of papers in both categories during the period between 2012 and 2022. While papers on the learner context and those on user context are relatively comparable in numbers before 2017, the trend for papers on a learner context slightly decreases afterwards, while the number of those discussing a user context relatively increases in comparison. This fact does not necessarily mean that learning contexts became less interesting for researchers, but rather shows the generalization aspect of modeling a user context in the recent years, whether for learning applications or other ones.

While a general user context may sound like an appealing concept, we further investigate those papers in terms of their compatibility with- and dependence on the pedagogical requirements for learning.

Our analysis shows that only 1.6% of those papers have any consideration of pedagogy or learning theories. With this result, one can clearly notice that the increase of research on contextualized solutions does not mean that the resulting systems are well integrable or implementable in a meaningful learning scenario, even if the algorithms were developed for general purposes that, theoretically, include learning application. This observation reveals a gap in the literature between context-aware algorithms that are designed based on pedagogical foundations, and those that are designed only based on technical considerations and artificial intelligence models.

## 4.1. Contextualization domains of application

While not limited to general-purpose contextualization, the tendency to use technical foundations for selecting context factors has also been visible in the literature that focuses on learner contexts. To analyze this aspect, we investigated the fields of application which were covered by the surveyed papers. Figure 4 shows the different fields that authors have focused on in their proposed context-aware approaches. As one can see from the application domain distribution, two fields are dominant in the literature: recommender systems and mobile learning (M-learning). Both domains form more than 31% of the surveyed papers. Following those domains are ubiquitous learning (U-learning) and adaptive e-learning. Remaining implementation fields are spread over 38% of the papers. Those included authoring tools, game-based learning, blended learning (B-learning), and adaptive e-learning, among others. Other papers have not addressed a specific application area, and therefore were included in a "General" category, covering 11.7% of the papers. The general category is not visualized in Figure 4 to enhance its readability.

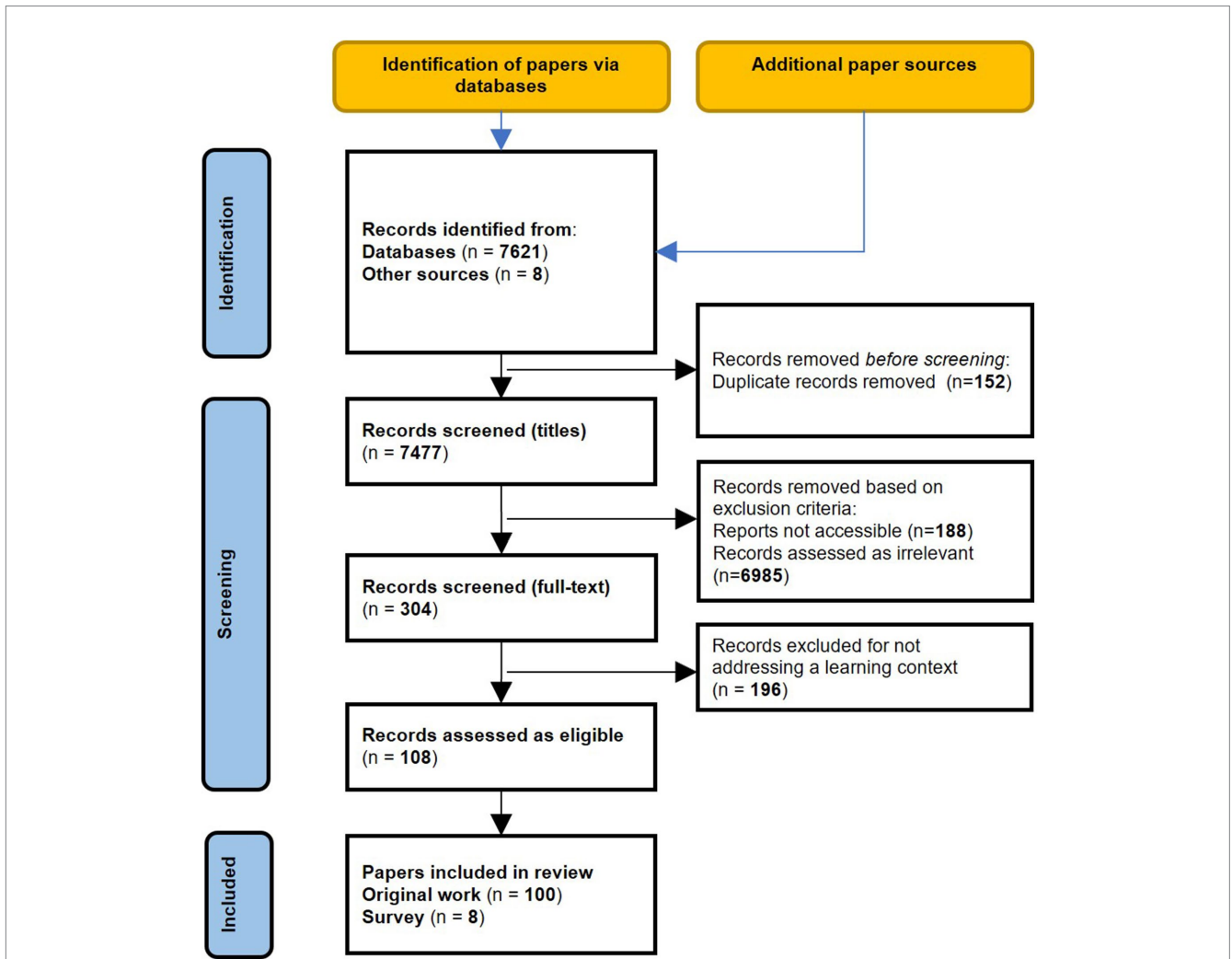

FIGURE 2
Search, screening, and selection process of the papers included in the review.

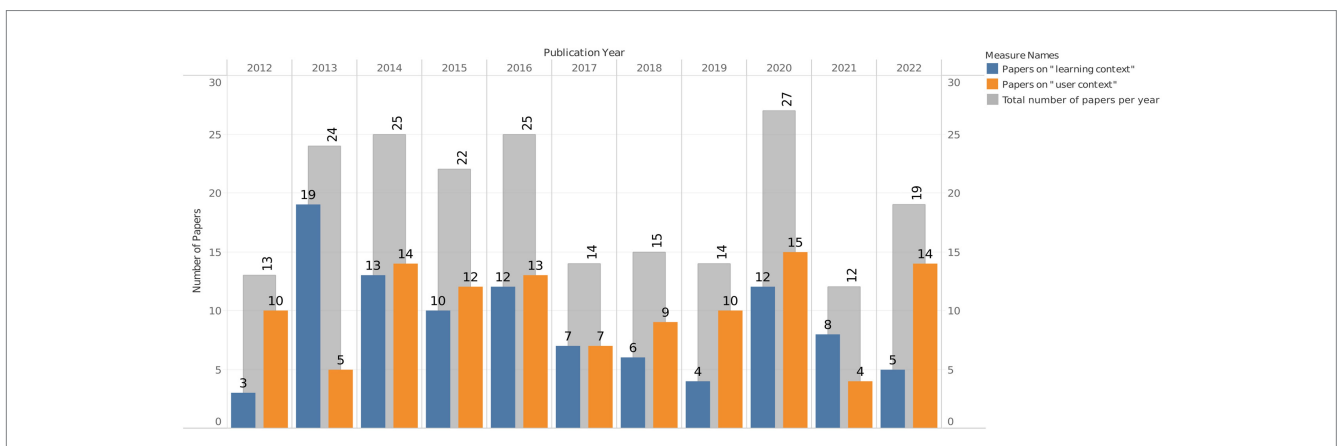

FIGURE 3
Number of papers covering "user context" (orange) and "learning context" (blue) as parts of the total paper sum per year (grey).

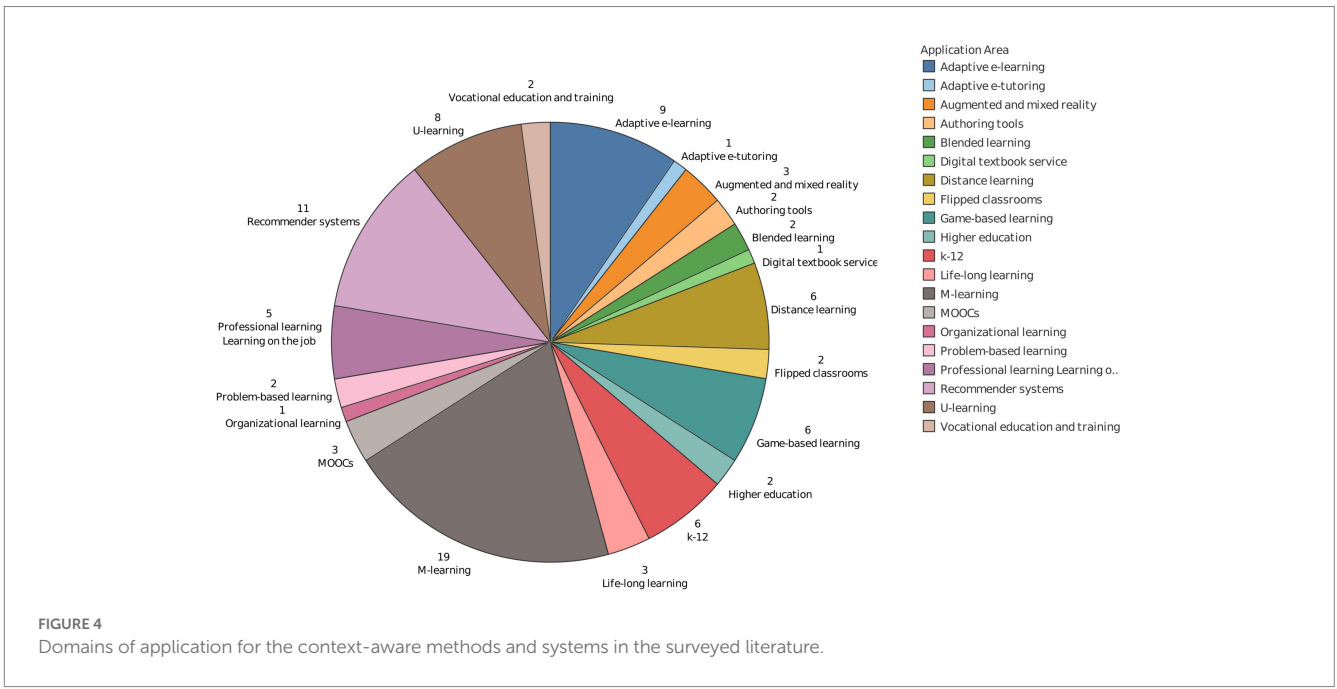

FIGURE 4
Domains of application for the context-aware methods and systems in the surveyed literature.

In the domain of recommender systems, Niemann and Wolpers (2013) develop content recommendation system in TEL, utilizing learning objects on web portals to find their semantic similarity. Shcherbachenko and Nowakowski (2018) utilize user context to enhance the accuracy of learning recommendation systems. The authors propose an architecture for a context-aware e-learning system based on the analysis of the means of creating recommendation systems for m-learning.

In the domain of m-learning, the growth of mobile device use by students allowed utilizing integrated sensors to capture further contextual data about the user. Mobile devices have been recognized as promising technological means that are able to facilitate teaching and learning for individual learners (Gómez et al., 2014; Sotsenko et al., 2016). This means an increased potential for personalization of the content delivered through the mobile device, based on the learner's characteristics and situation (Gómez et al., 2014). This argument has been supported in the literature by Alnuaim et al. (2016) and Ryu and Parsons (2008) who argue that mobile learning can enhance the learning process of students and help in fulfilling their personal needs due to the information it provides about the learners. Moebert et al. (2016) propose a generalized approach integrating context in mobile learning applications, to provide a contextualization framework that can be implemented in different domains and by different learning scenarios.

The support the contextualization offers is not limited to the learners, but also extends to support the content creator on educational authoring tools, such as in the work of Gallego et al. (2013), who propose a context-aware recommender system that suggests educational resources for the content creator when they are adding new learning objects on the authoring tool. The authors point out that the context they are capturing and using is not only one of the learners, but also one of the learning materials. This educational context can then include information about the material such as the language and the target age, which allows linking this material to a certain user's context.

A part of the publications in the surveyed literature includes original work that focuses on the context only from a data-driven perspective. In other words, it relies on the data from, e.g., mobile devices, to represent the learner's context. Other publications show pedagogically informed approaches for handling the user's context. Those consider requirements from pedagogy, and rules from learning theories, to define and use the learner's context. In order to differentiate both contextualization approaches, we investigate context dependence on learning theories, as well as its dimensions in the following section. Then, we further introduce categorization approaches to formulate their differences and similarities.

## 4.2. Learning theories in context-aware approaches

Considering the difference and overlaps between the definition of context in the surveyed literature, see section 2.4, we have found that the majority of surveyed literature falls into one of two categories: (1) one that pointed out, explicitly, the need for building learning-personalization solutions on pedagogical foundations, such as in Peña de Carrillo and Choquet (2013), Soualah-Alila et al. (2013), Gómez et al. (2014), or (2) one that implicitly addressed the pedagogical aspects of learning by focusing on contextual factors that correspond to one or more learning theories, such as the work of Musumba and Wario (2019) who propose and architecture of adaptive e-learning, which builds the user and material profiles taking into consideration the learner's decision-making process and the social aspects of learning, thus reflecting, implicitly, on problem-based learning, student-centered learning, and social constructivism theory. Figure 5 shows the main learning theories that have been addressed in the surveyed literature.

Bougsiaa (2016) and Zimmerman et al. (2016) both utilize augmented reality (AR) as a ubiquitous learning tool to provide a contextual learning experience to students. Bougsiaa argues, based on

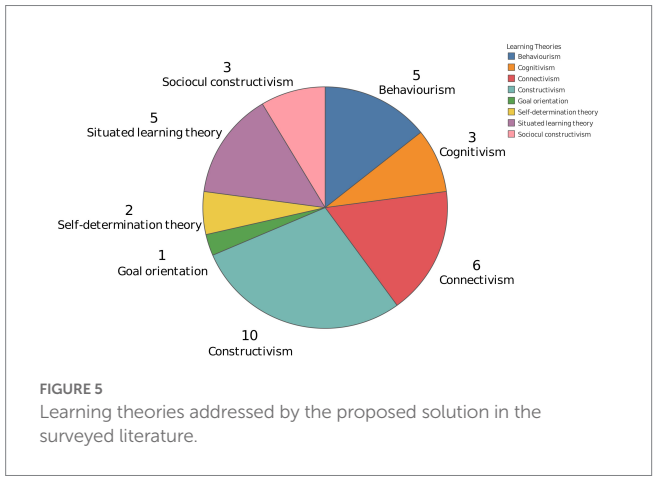

FIGURE 5
Learning theories addressed by the proposed solution in the surveyed literature.

Dunleavy et al. (2009) that AR aligns well with both situated learning and constructivism theories. This is because the learner is situated in a real-world physical environment, as well as a potential social setting (depending on the AP application), which offers the learners guidance and facilitates participation in a meta-cognitive learning process (Bougsiaa, 2016). Zimmermann et al. argue that AR offers a learning setting that supports disciplinary thinking, by providing just-in-time information from the surrounding environment, as well as incorporating scaffolding in the learning process (Zimmerman et al., 2016).

Gamified learning has also been proposed as a means for realizing self-determination theory, as the works of Shi et al. (2014), and Hwang and Chang (2020) show. The realization takes place through supporting the learner's autonomy in decision-making during the game, and their competence toward controlling the learning outcomes, and thus mastery of the learning content. This approach was also meant to increase the learner's motivation in a social e-learning environment (Shi et al., 2014). This approach furthermore reflects back on the goal orientation theory, by utilizing a ubiquitous computing environment to enhance learning efficiency through increased motivation in the learning process (Chiou and Tseng, 2012). In that regard, Chiou et al. have built on the theoretical concepts of goal orientation to develop a support system for a ubiquitous learning environment to offer navigation guidance for the learners among the existing learning content (Chiou and Tseng, 2012).

## 4.3. Context dimensions

The definition of learning context indicates that the situation in which learning takes place, describes not only the learner's situation, but also the environment and the interaction between them. This definition sets the first foundation for categorizing the learning context into different dimensions, which are here: the user, the environment, and the interaction means.

Throughout the research of learning contextualization, scholars have extended those dimensions to include finely granulated ones, which represent smaller or more specific groups of context factors. The bases for selecting a certain dimension, or group of dimensions, varied from one researcher to another. While some literature specifies dimensions of contextual factors corresponding to different sensory information that, e.g., a mobile device can collect, other literature defines dimensions that detail the learning pedagogical setting and thus separate, e.g., the interaction with a teacher from the interaction with school administration staff.

An early thorough investigation of the contextual dimensions has been introduced in 2012 by Verbert et al. (2012), where the authors identified eight dimensions of the learning context in TEL: (1) Computing, which includes the software and hardware specification, network characteristics, etc. (2) Location, including the quantitative GPS readings, or the qualitative values, such as home or school. (3) Physical conditions, which describe the surrounding of the user, such as the lighting and sound conditions. (4) Time, which may refer to the point of accessing the learning material, or the duration of learning. (5) Activity, which refers to the actions or tasks the user conducts in the learning process. Those are usually captured as events on, e.g., a learning platform during a learning session. (6) Resource, which describes the learning resources, in terms of their technical aspects (video, audio, text, etc.) or the metadata associated with them, such as resource annotations. (7) User, which describes the basic information about them, in addition to their knowledge, interests, and goals. (8) Social relations, which include the interactions with peers, colleagues, or educators in the learning setting.

Cerinšek et al. (2013) explore the contextual enrichment of a competence model. They preview TEL from a business- and organization-oriented perspective, which leads the authors to recognize four context dimensions: individual, organizational, knowledge, and environmental. While Verbert et al. address the technical and pedagogical perspectives of the contextual factors, Cerinšek et al. seem to focus on qualitative factors within each of the previous dimensions. For example, the individual dimension includes information about the learner, such as their ability to multi-task, or ability to acre knowledge. Organizational factors include the organization's ethics, social responsibility, and multidisciplinary. Knowledge dimension describes legislation factors, as well as business models, resource management and planning. Environment factors include the funds available, community stakeholders, market trends and climate changes, among others.

Sudhana et al. (2013) categorize the context factors into two dimensions: static and dynamic. Static dimension includes the user's personal details and environment factors, while the dynamic dimension includes the user preferences. This categorization is based on the contextual information acquisition. Here, the environment dimension includes the locations and time as factors acquired from the user profile and interaction with the system. The authors also categorize the learning context from a learner's perspective into three dimensions: (1) the learner's situation, which includes the environment of learning, such as the device. Location and time. (2) learning domain, representing the details about the subject, or learning area. (3) learner's activity, which includes the learning approach and the interaction events.

Yoo et al. (2013) consider the contextualization process from the perspective of a recommendation system algorithm. The authors classify the context factors under three dimensions: (1) The source, which represents the source of the recommendation, i.e., the recommender system itself. (2) The message, which is the recommendation that is communicated toward the learner. (3) The receiver, which is the learner who gets the learning recommendation.

Mohammad et al. (2015) analyze the learning context in massive open online courses (MOOCs). The authors classify context dimensions in that domain as: course context, class context, and student context. Bicāns (2016) addresses the context in classrooms and intelligent tutoring systems, where the author recognizes five dimensions of context modeling. Those dimensions are related to the learning object, pedagogy, student, learning session, and the learning environment. Li (2016) addresses contextual factors in MOOCs within the dimensions: external context, learner's context, social interaction, instructional design, and delivery platform. The author arranges those dimensions morphogenetically, i.e., sequentially in regard to time, to describe the different contextual levels, in which the learners find themselves throughout the learning process. Hemmler and Ifenthaler (2022) also classify contextual factors as external and internal, with the former referring to the environment and setting of learning, and the latter to factors that are related to the learner's themselves. The authors further extend the context dimensions within these two categories covering fine-grained aspects of the learner, environment, and resource.

Moebert et al. (2016) classify context dimensions as: (1) Physical context that describes the surrounding environment. (2) Mobile context that includes the location and movement factors. (3) Situational context that represents the physical situation of the learner, e.g., their facial expressions or body gestures. (4) Scenario, which represents the learning progress, tasks, and the time required for those tasks. (5) Personal context that represents previous knowledge, motivation, preferences, social interactions, and expectations. (6) Technical context, which is the context of the infrastructure used for learning, such as available device types and the device specifications.

Huang et al. (2017) classify context into: time, location, task, device, social and environmental, defining each of those dimensions similarly to Verbert et al. (2012). The authors further propose a framework for the learner-characteristics that includes the learner's basic information, cognitive level, learning style, and interest preferences.

Macchia and Brézillon (2021) proposed three overarching dimensions, or frames, to map learning objects and pedagogical methods in the pedagogical training. Those frames are then further broken down to contextual dimensions that are similar to previous classifications. Proposed dimensions are: (1) Learner, which includes the contextual factors within the sub-classes of physical characteristics and state, socio-cultural ones, emotional, intellectual, motivational, learning objectives and learning profile. (2) Training, which represents the domain and includs the sub-classes of training objectives, operational objectives, global pedagogical objectives, and the session context. (3) Learning activities, which represent the learner's interaction and manipulation of the learning objects.

Lallemand and Koenig (2020) categorize context factors in the domain of user experience (UX) within six dimensions: (1) Physical, which describes the physical conditions in which the learning application is deployed. (2) Social, which describes the interaction between the user and other users. (3) Internal, which refers to the user's status, such as their motivation and expectations. (4) Technical, which refers to the technologies used and the user's experience with them. (5) Task, which in this classification scheme does not refer to the task description, but rather to the user's perception of the task, such as the focus they devote to the task, their control over it, distractions from it and the potential multitasking. (6) Temporal, which describes a range of time-related factors such as the time of interacting with the system, duration of the interaction, its frequency, or time pressure during the interaction.

# 5. Results and discussion

In this section, we discuss the results, patters, and lessons-learned from the literature. We point out categories in the literature, which were not explicitly addresses in previous works, by introducing two categorization schemes of the context factors. The first scheme points out the pedagogical and technical origins of context in the literature, especially when those two origins overlap or differ. The second scheme points out the different profiles those factors belong to, when used in a digital learning platform.

We support our findings with a deeper discussion on the observations we have made from the literature landscape between 2012 and 2022. We formulate four observations that highlight critical issues we identified from the literature analysis and provide an overview on potential research directions in the future, to bridge the identified gaps and promote more alignment between pedagogy, TEL, and PALE.

## 5.1. Context-factor origins

Based on the overlap and differences between learning theories and technological solutions in defining the learning context and developing the contextual learning approach, see section 2.4, we identify a categorization of the learning contextualization, which points out the origins of the proposed concepts and solutions. We have concluded three categories in the surveyed literature:

- *Pure pedagogical contextualization*: In which, concepts from learning theories are considered for defining the learning context and its indicators. Papers in this category usually present deep philosophical argumentation on the context meaning and its implication in terms of constructing knowledge from the learning process. Papers belonging to this category discuss the context on a high-level, and seldomly address how the context is being captured or measured in a real-world scenario. Examples of this category can be found in Negovan and Bogdan (2013), Opel and Brinda (2013), Pimmer et al. (2013), Tempelaar et al. (2013), Danes and Brewton (2014), Fancsali and Ritter (2014), Shi et al. (2014), and Buchmann (2022).
- *Pure technical contextualization*: This category represents the opposite side of the contextualization approaches, in which the definition of the context is solely based on the devices and technological solutions used to capture learners' data. Papers in this category focus on the means of measuring contextual data through sensors and technological observations, and therefore define the context of learning directly from that angel, as one can see in Chorfi et al. (2012), Gallego et al. (2013), Hwang (2013), Niemann and Wolpers (2013), Qiuyan et al. (2013), Ku (2014), Clarkes-Nias et al. (2015), Madhu Sudhana (2015), Baccari and Neji (2016), and Supic (2016).

- *Pedagogically informed technical contextualization*: Which is the category that includes technical implementations of contextualized learning, which are based on solid, explicate or implicit, foundations in learning theories and pedagogical requirements. Papers belonging to this category argue with the relations between certain learning technologies on the one hand, such as AR, recommendation systems, mobile devices, or learning games, and on the other hand, one or more specific learning theories that are realized through those technologies. Examples of publications in this category can be found in di Mascio et al. (2013), Hampel and Arcos (2013), Knobelsdorf and Tenenberg (2013), Peña de Carrillo and Choquet (2013), Reynolds and Chiu (2013), Brézillon (2014), Colman et al. (2014), Edmonds (2014), Forissier et al. (2014), Mowafi et al. (2014), Parker and Hollister (2014), Ramakrishnan et al. (2014), Borgnakke (2015), Evans (2015), Mohammad et al. (2015), Alnuaim et al. (2016), Moebert et al. (2016), Ciordas-Hertel (2020, 2022), and Herrero-Martín et al. (2022). In the survey time span, papers in this category formed 58% of the surveyed literature, in comparison to 23% purely pedagogical approaches, and 19% purely technical ones.

Table 2 shows the distribution of surveyed papers over the context-origin categories.

## 5.2. Profile-based context dimensions

Surveyed approaches have proposed a range of dimensions, which intersect and differ on multiple levels. It can be concluded that the domain of application (Zbick et al., 2016), the tools and infrastructure of the learning (Hwang, 2013), and the reliance on technical or pedagogical foundations for the learning (Macchia and Brézillon, 2021), are all factors that influence the adoption of a certain dimension, or set of dimensions, over the other ones.

On a learning platform, information about learners, learning resources and the learning environment is usually included in the form of profiles. A learner has a unique profile that includes information about their age, interests, learning goal, etc. This is also the case with learning resources, whose profiles include information about the type of the resource, its length, etc. Information about the environment, such as the lighting and noise levels, is usually measured and included in a temporary "learning session" profile. While previous classification schemes address context factors within different learning settings, one classification approach that is still missing is one that addresses how learning context factors are practically handled on learning platforms. A learner profile, for example, covers multiple dimensions about the learner/user from the schemes above, including the "Physical," "Task," "Internal," "Situational," and "Personal" classes. To address this issue from a practical perspective, which takes into consideration the different profiles that context factors belong to, on a learning platform, we propose a new scheme for categorizing contextual factors within three dimensions: (1) Learner, (2) Educational resource, and (3) Environment. The latter dimension is broken down to sub-classes that describe technical, dynamically changing factors, such as the noise and light levels, as well as pedagogical ones such as the organizational structure, teacher assessment, or parent and teacher involvement. Figure 6 illustrates the proposed dimensions and their sub-classes. We try through this classification to accommodate the majority of frequently used contextual factors in the literature, from pedagogical and technical points of view. We define the dimensional structure on multiple levels of granularity, which is then adaptable to domain-specific requirements, and the technical implementation requirements, while being at the same time aligned with the common practices of learner and resource profiling on learning recommendation platforms.

## 5.3. Context capturing and contextual factors

In TEL, contextualizing the learning process is only possible if the learning context is captured and represented accurately in the technological solution. Based on the different definitions of learning context, different factors are defined to represent it. These factors are meant to provide a quantitative representation of otherwise qualitative definitions of a contextual aspect of the learning. For example, the social context of a learner is a contextual aspect originated from constructivism theory, which can only be implemented in a learning recommender system if there is a quantitative value that the recommendation algorithm can use for calculating the top-n recommendations. Such quantitative values might be: the user-profile-IDs that are connected to the current learner's profile; or the number of shared courses that are attended by the current learner and another member of a social group to which the learner belongs. Without these types of quantitative values, the recommender cannot assess or utilize the indication of the social factor for generating a learning suggestion.

Reflecting on the definitions of a learning context in the three categories in section 5.1, one can notice that the quantitative nature of the technical algorithms that are used for learning contextualization is sometimes challenged by the pedagogical definition of that context. For instance, legal or psychological aspects of the context (Buchmann, 2022) are not simple data structures or parameter-values that can be easily measured, e.g., using a sensor. They are also complex for an evaluation done by a human factor, e.g., through a survey, since they require specialized experience and knowledge. In contrast, a contextual factor like the location, which can be selected from a list, or captured from a mobile's global positioning system (GPS), shows the considerable difference in the difficulty of capturing the value of a contextual factor, when compared to, e.g., the legal context of learning, or the socio-economic status of the learner. Such challenge, and other similar limitations, led some scholars, such as Taylor et al. (2021), to argue that bringing the pedagogy, which recognizes an individual level of personalization, up to scale requires technologies that are not yet available. That, however, does not mean that personalization based on an individual learner's context is not possible. What it means, as we interpret it, is twofold: (1) contextualization technologies and algorithms should address the pedagogical aspects, such as the legal factor, to the best of their ability, while addressing their limitations in capturing or representing those factors, rather than avoiding them all together as in the majority of technical solutions. (2) if contextualization approaches seek to be pedagogy-compatible, to bring pedagogy to a real scale, qualitative contextual factors should be researched more to find suitable quantitative representations for Abu-Rasheed et al.    10.3389/feduc.2023.1210968TABLE 2 Frequent contextual factors, their context-origin, profiling categories, and examples of papers addressing them in the surveyed literature.

| Factor | Contextualization foundation | | Context profiling category | | | | Addressed in |
|---|---|---|---|---|---|---|---|
| | Pedagogical | Technical | Pedagogical+ Technical | User | Resource | Environment | |
| Location | | | • | • | | • | Alnuaim et al. (2016), Baccari and Neji (2016), Bougsiaa (2016), Siraj et al. (2018), Musumba and Wario (2019), Ciordas-Hertel (2020) |
| Time | | | • | • | | • | Sevkli and Abdulkarem (2015), Supic, (2016), Brik and Touahria (2020) |
| Previous knowledge | • | | | • | | | Bicāns (2016), Tarus et al. (2018), Hilkenmeier et al. (2021), Bicans and Grundspenkis (2017) |
| Peers / colleagues | | | • | | | • | Shcherbachenko and Nowakowski, (2018), Daoudi et al. (2020), Lallemand and Koenig (2020), Starr (2020), Dennen et al. (2018) |
| Activity | | | • | • | | | Bouihi and Bahaj (2017), Li et al. (2017), Nijenhuis-Voogt et al. (2018), Junianto and Wutsqa (2019) |
| Interests | • | | | • | | | Thüs et al. (2015), Zimmerman et al. (2016), Mayeku and Hogrefe (2017), Brik and Touahria (2020) |
| Instructional strategies | • | | | | • | • | Caponera and Losito (2016), Kuger et al. (2016), Zhou et al. (2021), Buchmann (2022) |
| Motivation | • | | | • | | | Li (2016), Lallemand and Koenig (2020), Mutambik et al. (2020), Starr (2020) |
| Noise level | | • | | | | • | Zheng et al. (2019), Ciordas-Hertel (2020), Lallemand and Koenig (2020), Zaguia et al. (2021), Ciordas-Hertel et al. (2022) |
| Psychological state | • | | | • | | | Shcherbachenko and Nowakowski (2018), Daoudi et al. (2020), Ilkou et al. (2021), Buchmann (2022), Ciordas-Hertel et al. (2022) |
| Culture | • | | | | | • | Kuger et al. (2016), Bidarra and Rusman (2017), Nijenhuis-Voogt et al. (2018), Aldowah et al. (2019), Mutambik et al. (2020) |
| Ambient light | | • | | | | • | Huang et al. (2017), Mausz and Tavares (2017), Ciordas-Hertel (2020), Lallemand and Koenig (2020), Ciordas-Hertel et al. (2022) |
| Learning goal | | | • | • | • | | Bicāns (2016), Mayeku and Hogrefe (2017), Tarus et al. (2018), Daoudi et al. (2020), Lallemand and Koenig (2020) |
| Network bandwidth | | • | | | | • | Baccari and Neji (2016), Huang et al. (2017) |
| Socio-economic status | • | | | • | | • | Caponera and Losito (2016), Starr (2020), Zhou et al. (2021), Buchmann (2022) |
| Language | | | • | • | • | | Baccari and Neji (2016), Aldowah et al. (2019), Daoudi et al. (2020), Starr (2020) |
| Device interaction type | | • | | | | • | Alnuaim et al. (2016), Dennen et al. (2018), Shcherbachenko and Nowakowski (2018), Lallemand and Koenig (2020), Zaguia et al. (2021) |
| Organization | • | | | • | | • | Caponera and Losito (2016), Bidarra and Rusman (2017), Aldowah et al. (2019), Mutambik et al. (2020), Hilkenmeier et al. (2021) |
| Learning history | | | • | • | | | Bicans and Grundspenkis (2017), Huang et al. (2017), Musumba and Wario (2019), Chung (2020), Ilkou et al. (2021) |
| Weather conditions | | | • | | | • | Zheng et al. (2019), Lallemand and Koenig (2020) |
| Physical state | | | • | • | | | Daoudi et al. (2020), Lallemand and Koenig (2020), Macchia and Brézillon (2021) |
| Task | • | | | | • | • | Huang et al. (2017), Siraj et al. (2018), Hwang and Chang (2020) |
| Legal/Laws | • | | | | | • | Aldowah et al. (2019), Starr (2020), Buchmann (2022) |
| Teaching materials | | | • | | • | | Caponera and Losito (2016), Dennen et al. (2018), Musumba and Wario (2019), Nuankaew and Nuankaew (2019) |

(Continued)14

TABLE 2 (Continued)

| Factor | Contextualization foundation | | Context profiling category | | | | Addressed in |
|---|---|---|---|---|---|---|---|
| | Pedagogical | Technical | Pedagogical+ Technical | User | Resource | Environment | |
| Ambient temperature | | | • | | | • | Huang et al. (2017), Ciordas-Hertel et al. (2022) |
| Wireless communication | | • | | | | • | Huang et al. (2017), Siraj et al. (2018), Aldowah et al. (2019), Daoudi et al. (2020) |
| Preferences | | | • | • | | | Musumba and Wario (2019), Amasha et al. (2020), Canbaloğlu and Treur (2022), Ilkou et al. (2021) |
| System interactions | | | • | | | • | Zarrad and Zaguia (2015), Shcherbachenko and Nowakowski (2018), Tarus et al. (2018), Chung (2020) |
| Device ID | | • | | | | • | Baccari and Neji (2016), Bouihi and Bahaj (2017) |
| Age | | | • | • | | | Bouihi and Bahaj (2017), Daoudi et al. (2020), Wongchiranuwat et al. (2020) |
| Classroom condition | • | | | | | • | Caponera and Losito (2016), Zheng et al. (2019) |
| Battery charge | | • | | | | • | Siraj et al. (2018) |
| Screen size | | • | | | | • | Alnuaim et al. (2016), Siraj et al. (2018) |
| Ethics | • | | | • | | • | Kuger et al. (2016), Starr (2020) |
| Parental involvement | • | | | | | • | Caponera and Losito (2016), Zhou et al. (2021) |
| Movement | | • | | | | • | Shcherbachenko and Nowakowski (2018), Ciordas-Hertel et al. (2022) |
| Gender | • | | | • | | | Mutambik et al. (2020) |
| Teacher involvement | • | | | | | • | Aldowah et al. (2019), Herrero-Martín (2022) |
| Device type | | • | | | | • | Lallemand and Koenig (2020) |

them, which the algorithms can use. We argue that this effort lies on both pedagogical and technical experts equally since a correct solution is not feasible without a close collaboration between the two domains. In the surveyed literature, we only find traces of this concept in the work of Moebert et al. (2016), who classified the relevance of different contextual factors to a pre-defined list of educational settings (formalized, physical, collaborative, immersive, as well as teaching and learning). The authors also classify the measurement accuracy of those factors within each setting. They propose a framework for detecting, collecting, and utilizing contextual factors in adaptive mobile learning applications, focusing on the framework's adaptation to different learning settings, as different factors vary in their relevance to each setting and measurement accuracy within it.

Capturing a learning-context factor is dependent on the definition of that factor, the underlying algorithm that is using it, and the tools utilized in the learning task. To elaborate on this point, we use the "location" contextual factor as an example. Location may refer to the physical location of the learner, in terms of their geographical coordinates; and may refer to their virtual location, such as their location on a network of contented devices. Following each definition, a different approach is used to capture the learner's location, such as using a GPS reading for the former definition and the device's internet protocol (IP) address for the latter. On the other hand, algorithms and tools also play a role in determining the context capturing approach. A mobile learning personalization utilizes the sensors of a mobile device to read contextual information about the learner, while a recommendation algorithm that is implemented on a web-based learning platform utilizes the learner's profile to get contextual information about their context, which is usually done manually by the learner themselves when signing up for the learning platform, i.e., providing the city (location) in which they are located.

Another approach for capturing context factors, which are more qualitative in nature and mostly close to the pedagogical domain, is the use of surveys that are analyzed and evaluated by domain experts. An example is seen in the work of Caponera and Losito (2016) through capturing the socio-economic status of learners. The program for international student assessment (PISA) (Kuger et al., 2016) also captures a wide range of contextual factors that influence the student's learning. The study utilizes tests and surveys to collect data on the user performance and other aspects that may influence it, including the learner's context.

In the surveyed literature, we investigate a range of contextual factors, as well as their frequent use in the proposed solutions and approaches. Factors that we identify, and those identified in other valuable research, as in Caponera and Losito (2016), Kuger et al. (2016), Lallemand and Koenig (2020), Hilkenmeier et al. (2021), and Hemmler and Ifenthaler (2022), are addressed with different frequencies in the literature. Reasons for a factor to be a "commonly used" one are related to: (1) the origin of that factor, i.e., pedagogical, technical, or both. (2) the ease of capturing that factor. (3) factor's influence on the learning personalization. We illustrate the different frequencies in which most common factors appear in Figure 7. We place the contextual factor between pedagogical and technical origins as it is addressed by literature from both domains. In this sense, factors that are closer to the diagonal line are those that belong to the third category in section 5.1 (Pedagogically informed technical contextualization). The size of the factor refers to how frequently it has been addressed in the literature. Here, we consider that the contextual factor addressed in the paper if the authors explicitly mention that factor as a means of representing the learning contexts, even if the authors did not use that factor in their proposed solution. This is because the selection of factors in each solution is subject to other influences, such as the domain of application, the learning tools and the algorithms use. We include in Figure 7 only those factors that have a frequency $f \geq 3$ due to the large number of contextual factors that are less frequent. A list of those factors and their classifications, as well as papers that addressed them, is provided in Table 2.

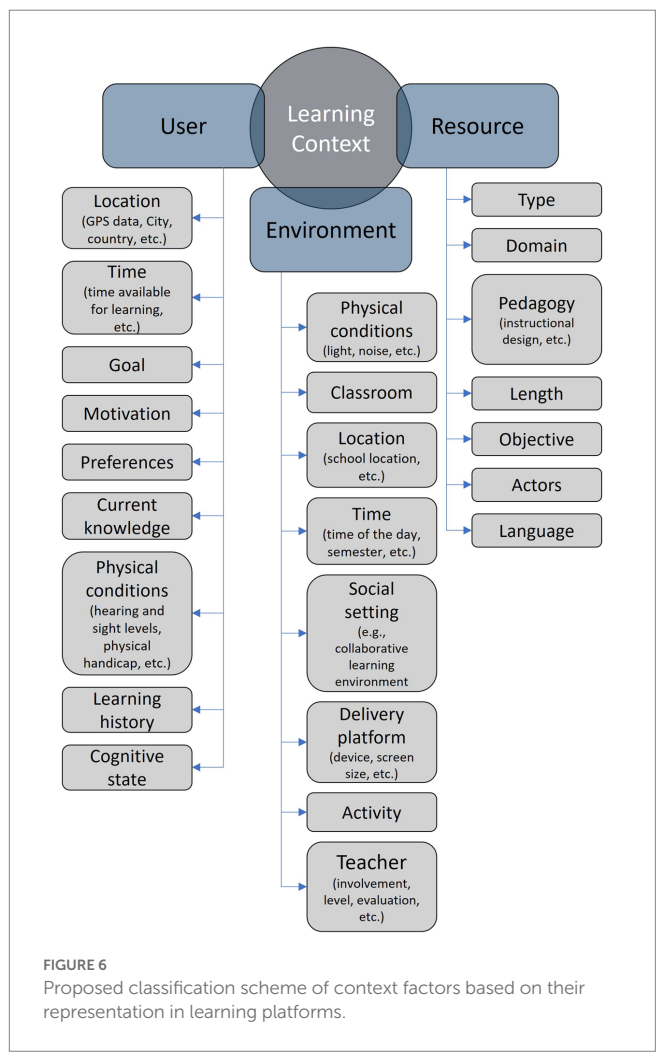

FIGURE 6
Proposed classification scheme of context factors based on their representation in learning platforms.

It is noticeable from Figure 7 that location and time are among the most frequently addressed factors in the literature. We argue that the reason can be the strong pedagogical foundation of those factors, in terms of their influence on learning, as well as the feasibility of capturing them in most of the technical solutions used in learning personalization algorithms. Other factors, such as the socio-economic ones, are commonly used in pedagogical approaches, since they have proven their influence on the learner's performance (Caponera and Losito, 2016). However, they are considerably less frequent than other factors, which may be a result of the difficulty in measuring them in a technical solution.

Similar to the location factor and the differences in its actual meaning and definition, other contextual factors that we identified in the literature experience a variety of definitions. Time as a contextual factor may refer to the time of the day when the learning is taking place, or the time of the year, e.g., season. Time exceeds referring to a single point to refer to the duration of learning or interacting with the learning tool. Therefore, some literature defines location and time as context dimensions, rather than context factors. In other words, they can refer to a group of spatio-temporal factors, instead of being only two individual ones, due to the variety of definitions they can take. The categorization of context factors within contextual dimensions has been followed in the literature using a multitude of dimension considerations. In the following section, we investigate the different dimensions identified in the literature for the contextual factors.

The analysis of the surveyed literature has led to a set of observations and conclusions about the state-of-the-art research on learning context. In the following sub-sections, we summarize those observations objectively, to point out key findings, patterns and limitations identified in the literature. Furthermore, we point out the challenges and open research topics that are associated with those observations and findings.

## 5.4. Observation 1: learning context between pedagogy and technology

Our review results show conceptual and practical gaps between the pedagogical and technical domains when it comes to defining a learning context and capturing its factors. Context definitions from the lingual, pedagogical or technical points of view appear to be similar to each other. However, the interpretations of the context meaning take different directions in the practical implementation. It is clear from the literature on context factors and learning contextualization methods that the pedagogical research focuses on high-level, conceptual, and sometimes abstract, aspects of the learning context. For instance, scaffolding learning points out the role of teacher intervention in supporting the learning process. It does not, however, define how the intervention takes place, how it can be described, or what its limitations are. Another example can be seen in defining the learner context in a vocational education setting using factors such as the learner's "ability to acquire knowledge," which is an abstract consideration that may face the challenge of defining what an "ability" is even before defining what the contextual factor represents. We argue that this tendency to utilize philosophical abstraction for describing the learning context in learning theories and pedagogical science takes place naturally for two reasons: (1) the cognitive nature of the domain itself, which requires the abstraction and conceptualization levels to define the context as a construct. (2) the implementation model in learning theories and pedagogical science, which naturally assigns the task of interpreting the fine-grained details on considering the learning context in a real-live implementation. This human factor is usually represented by the educators, who carry out the actual realization of a pedagogical approach, e.g., in a classroom.

Technical definitions of the learning context also have a clear tendency toward purely practical realizations of the context, by focusing on quantifiable and measurable factors that can describe the learning. Those factors are also bound to the technologies that measure them, such as considering the location factor virtually, in the form of a node of a network of devices, represented by a numerical IP address value. Although this concept seems to be an intelligent realization of the pedagogical concepts of learning context, which address the factors of connectivity to other learners on the network, e.g., from the connectivism and social constructivism theories, we have observed that the majority of the technical solutions do not address this pedagogical foundation. Yet, they continue to use the IP address as a virtual location factor, influenced by the increased accuracy it results in, when implemented in a social recommendation system, for example. In other words, the underlying reason for adopting a certain technical factor to represent the learning context in a technological solution, is not always that this factor serves a founding pedagogical goal, but rather because that factor is either easy to measure or has proven its usefulness in other technical algorithmic implementations that may be outside the learning or educational domains. The utilization of such simplified measures is, of course, not only because they are easy to implement, but because of an implicit assumption or expectation from the authors that the simple version of a contextual measure, e.g., IP address, still has a sufficient correlation to the targeted context to be represented, e.g., social setting. This assumption is, however, seldom justified from a pedagogical perspective.

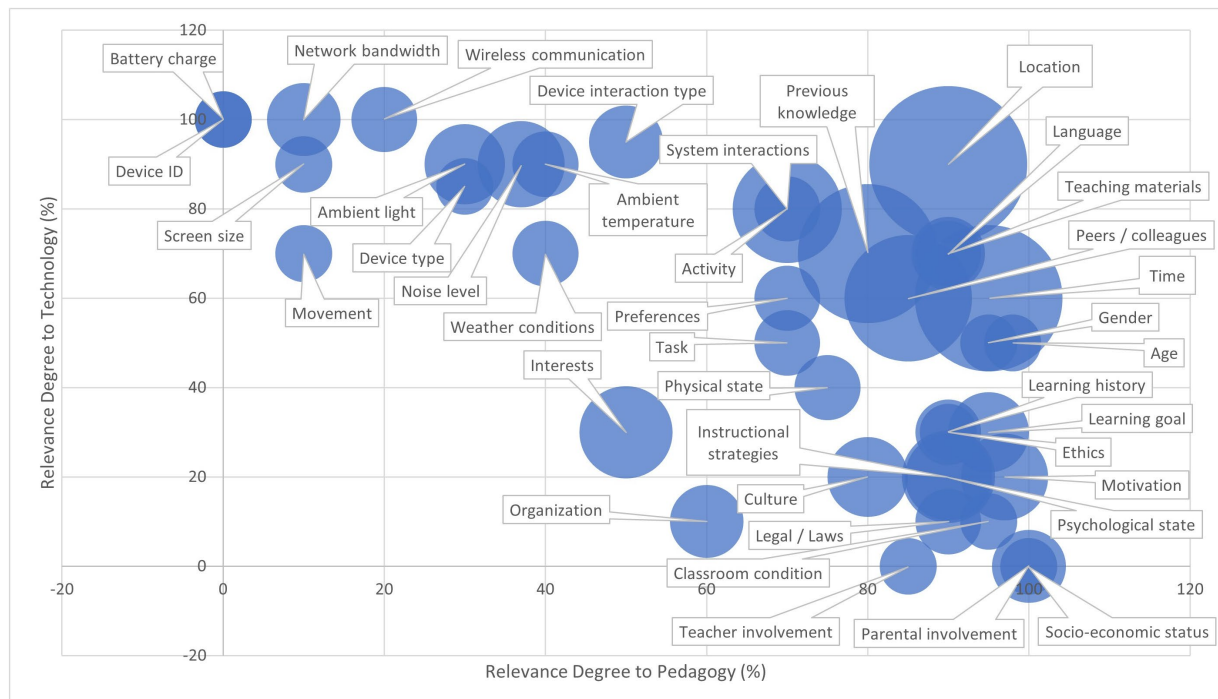

FIGURE 7
Mapping frequent contextual factors based on their relevance to pedagogy and technology. Size of the factor reflects its frequency in the surveyed literature.

Several simplified factors, such as "reviews" or "number of likes" that are implemented in context-aware educational recommendation systems have their origins in product recommendation algorithms, e.g., book or movie recommendations. We observed that no authors provide a clear pedagogical argument about using similar factors for learning recommendations. While it is not clear to us when this approach started in the literature, the researcher will not miss that approach when observing the evaluation strategies of context-aware recommendation systems, which mostly rely on large datasets that seldom include pedagogically defied context factors.

It is important to point out that many technical solutions for learning personalization are explicitly or implicitly based on pedagogical foundations. This has been observed in the surveyed literature in section 5.1. However, there is a gap in defining the context and projecting and implementing each contextual factor. This is observable within the lack of a standardized or unified approache to define each context factor and each context dimension in technical implementations.

## 5.5. Observation 2: context dimensions as domain-specific or standardized schemas

The surveyed literature has revealed a considerable variation in classifying contextual factors. Classification schemes and approaches were based on a variety of considerations, such as the pedagogical settings, implementation infrastructures, and the domain-specific requirements. Moreover, classification schemes used, on several occasions, dimensions identified in other schemes but wither with a different definition or different level of granularity. For example, location is considered a concrete factor in some schemes, which belongs to the learner or the environment dimensions, while in other schemes it is considered a dimension, representing a group of factors related to the physical or virtual locations.

The need for a standard classification of context factor dimensions is key to enable aligning context-aware solutions with each other. Furthermore, the reusability of a certain contextual dimension from one system in another is only possible if its definition and factors of that dimension are the same in both systems. Definition uniqueness, reusability and interpretability are among the key elements of implementing FAIR principles for scientific data management (Wilkinson et al., 2016), which offer important guidelines for implementing the corresponding context-aware learning personalization, among other systems.

The lack of standards for describing the context factors and dimensions is not a new challenge. This issue has been addressed in Verbert et al. (2012), and it seems to remain unsolved till now in the state-of-the-art, despite several attempts to propose generalizable schemes of the context dimensions, such as in the work of Zbick et al. (2016), Moebert et al. (2016), and Li (2016). One of the solutions that were proposed for this challenge, is to adopt a semantic-web-based approach for defining standardized vocabularies for context factors and dimensions. Utilizing a standard definition of, e.g., time or location, which has been described in detail in a semantic web schema, such as (Schema.org) from the World Wide Web Consortium (W3C), may enable a global reusability of the factor of dimension among context-aware solutions.

## 5.6. Observation 3: context capturing and quantification

The difference in defining a contextual factor between the technical and pedagogical domains is not the only challenge identified for methods used to capture and represent that. The ability to represent or model a contextual factor such as "Ethics" is challenged by the abstract nature of that factor. We observed a general difficulty in capturing and quantifying several contextual factors originating from pedagogical approaches and learning theories. This difficulty results from: (1) the lack of clear methods to quantify a qualitative factor, such as "the involvement of parents" in learning. (2) the lack of guarantee that the quantification approach preserves the pedagogical meaningfulness of a contextual factor. For example, if a qualitative factor like the "interaction with the teacher" is quantified by counting the number of conversations the student had with the teacher, would this quantification reflect the real meaning of teacher interaction as intended by scaffolding learning? The answer to this question is only possible through pedagogical experts. Therefore, in situations where context-aware algorithms are designed and implemented by technical experts and software developers, there is no guarantee that the quantification method is based on a correct interpretation of the pedagogical requirements. This observation highlights the importance of collaborations between experts from both domains, when designing and implementing personalization algorithms in general, and context-aware algorithms in particular.

## 5.7. Observation 4: explicit and implicit learning context consideration

In the search phase for literature on learning context definitions and factors, we came across a number of publications, which only implicitly address learning context factors, e.g., in the design of a learner profile, without explicitly naming those factors as "learning context factors." This is also the case for publications that describe the learning context itself, but only indirectly, such as describing the situation of the learner or the learning setting, without addressing it explicitly as a "learning context." The effect of this is the difficulty in finding the complete spectrum of literature about learning context using explicit search terms. It is, in fact, the reason we had to include several additional papers in this review during the paper identification phase. This translates into a direct limitation of surveys of this kind.

We observe that the majority of papers that address the learning context implicitly are related to topics that require user- or learning resource profiling. While application domains, technical solutions, and points of interest of those papers differ, they still address a specific representation, or a model, of the learner. That representation includes then metadata, which describes one or more contextual factors. Examples of this literature can be found in Ciloglugil and Inceoglu (2018), Barria-Pineda et al. (2019), Chimalakonda and Nori (2020), Ilkou et al. (2021), and Pal et al. (2021), where the authors utilize ontological classes to describe the situation of the learner or the learning material, without addressing those as contextual classes.

One potential reason for this observation is related to the point of having no standard definition of context or contextual factors, which can be adopted and followed in the research on user and resource modeling. To that end, working toward a standard description of learning context, and frameworks that define clear roadmaps to considering its factors in pedagogical and technical methods, seems to be a field of research that still requires more investigation and attention. Here, it is important to address the fundamental issue with "modeling" as a process, which is the fact that models are always imperfect and only describe reality with a controlled compromise of accuracy. In this sense, the technical implementation of learning context factors can always follow one of many implementation approaches, where each one is still imperfect as a general approach, but can still be optimal, or at least effective, in modeling a specific, well-limited, learning scenario. As such, having a database or semantic web2.0-based approach to listing factor implementations would have a great potential to add the assumptions and limitations of the modeling, as well as the goal of that specific implementation. With that in mind, there will not be a single implementation for each factor, but rather a limited range of implementations, which "sufficiently" cover the factor meaning in all its relevant scenarios.

Observation 6.4 objectively addresses a limitation of our survey, which is related to the literature coverage and finding all potential papers that address the contextual factors. Our search terms were selected to the best of our knowledge for collecting papers in multiple domains that are relevant to the learning context and contextual factors. However, other publications that implicitly address context factors may still be found. Therefore, a further extension of the literature coverage and knowledge base of identified papers holds the potential to include additional resources within the analysis we present in this review.

# 6. Conclusion

In this article, we surveyed the state-of-the-art learning context definitions and contextual indicators in personalized and adaptive learning and context-aware recommendation systems. Our survey covers the literature during the period between 2012 and 2022, which addressed the learning context, its dimensions, and factors. A systematic approach was followed to search and select the final set of publications to be reviewed. We identified 108 publications as relevant for the review, from three main repositories Google Scholar, Springer Link, ACM Digital Library, EbscoHost and LearnTechLib.

We analyzed the selected relevant publications from multiple perspectives, which aim to investigate: (1) the definitions of learning context in pedagogy and technology. (2) the different categories, to which context factors and indicators belong. (3) the pedagogical foundations of the context in the technological development of personalization algorithms. (4) the alignment and differences between pedagogy and technology in utilizing context factors. We also introduced two classification schemes for contextual factors, based on their foundation in pedagogy or technology, and based on their representation in learning platforms. Then, we concluded the survey with a thorough discussion of the results and observations we have made from the literature, which furthermore represent chances for a further extension and improvement of the field.

Our findings in this survey point out several gaps in the literature on learning contextualization. The lack of a standard definition of the learning context, and the resulting differences between pedagogical and technological considerations of its factors, have a clear effect on the methods implemented in both domains to capture this content and use it for personalizing the learning offerings. Q uantifying contextual factors that have a more qualitative nature, such as "ethics" or "laws" also reveal a challenge for the technical adoption of these factors in quantitative algorithms. The dependency on purely technical factors that result from sensor readings, such as the "device orientation" required more foundation in learning theories, to enable enhancing the learning personalization from a solid pedagogical perspective. Further research in these directions is recommended, as the lack of comprehensive solutions has been observed in this review of the surveyed literature.

## Author contributions

HA-R, CW, and MF: study objectives and validation. HA-R and CW: study design, research questions, search strategy implementation, identifying relevant studies, assessment of extracted-data quality, statistical analysis, visualization, and supervision. HA-R: data retrieval from literature and writing the original draft. CW: proof-reading and writing extension. HA-R and MF: project administration. All authors contributed to the article and approved the submitted version.

## Funding

This research has been partially funded by the INVITE call by the ministry of education and research BMBF and the federal institute for vocational education and training BIBB as part of the WBsmart research project (grant ID: 21INVI21/810305313807), Germany; as well as the ERASMUS+ Key Action 204 Higher Education project OSCAR (grant agreement ID: 2020-1-DE01- KA203-005713).

## Conflict of interest

The authors declare that the research was conducted in the absence of any commercial or financial relationships that could be construed as a potential conflict of interest.

## Publisher's note

All claims expressed in this article are solely those of the authors and do not necessarily represent those of their affiliated organizations, or those of the publisher, the editors and the reviewers. Any product that may be evaluated in this article, or claim that may be made by its manufacturer, is not guaranteed or endorsed by the publisher.